\documentclass[twocolumn,showpacs,preprintnumbers,amsmath,amssymb]{revtex4}

\usepackage{graphicx}
\usepackage{dcolumn}
\usepackage{bm}

\begin{document}
\preprint{APS/123-QED}

\title{Phonon laser action in a tunable, two-level photonic molecule}
\author{Ivan S. Grudinin}
\author{O. Painter}
\author{Kerry J. Vahala}
\email{vahala@caltech.edu}%
\affiliation{Department of Applied Physics, California Institute of Technology}
\affiliation{1200 E. California blvd, Pasadena, CA 91125, USA}

\date{\today}

\begin{abstract}
The phonon analog of an optical laser has long been a subject of interest. We demonstrate a compound microcavity system, coupled to a radio-frequency mechanical mode,  that operates in close analogy to a two-level laser system. An inversion produces gain, causing phonon laser action above a pump power threshold of around 50\,$\mu$W.  The device features a continuously tunable, gain spectrum to selectively amplify mechanical modes from radio frequency to microwave rates.   Viewed as a Brillouin process, the system accesses a regime in which the phonon plays what has traditionally been the role of the Stokes wave. For this reason, it should also be possible to controllably switch between phonon and photon laser regimes.  Cooling of the mechanical mode is also possible. \end{abstract}

\pacs{Valid PACS appear here}
\maketitle

$\it Introduction. \rm$ --- The possibility of phonon laser action has been considered in a wide range of physical systems including ions\cite{Wallen}, semiconductors\cite{Liu:2003cr}, nanomechanics\cite{Bargatin:2003zk}, nanomagnets\cite{Chudnovsky:2004pd} and others\cite{Khurgin}. Observations of phonon stimulated emission have been reported in cryogenic $\rm Al_2O_3$:$\rm Cr^{3+}$ \cite{tucker, Hu, Fokker} and $\rm Al_2O_3$:$\rm V^{4+}$ \cite{Bron} as well as semiconductor superlattices\cite{Kent:2006rw}. However, only recently has phonon laser action been reported using a harmonically-bound magnesium ion\cite{Vahalaion}.  Here, using a compound microcavity system,  a phonon laser that operates in close analogy to a two-level laser system is demonstrated. The approach uses intermodal coupling induced by radiation pressure\cite{braginsky01}, and can also provide a spectrally selective means to detect phonons. Moreover, there is currently great interest in optomechanical cooling\cite{kippenbergscience08} and evidence of intermodal cooling is observed.

Compound, microcavity systems are sometimes called photonic molecules\cite{Bayer_Molecule, Barnes_3D_Molecule, Nakagawa_Disk}; and herein this is a particularly apt interpretation.  Hybridized orbitals of an electronic system are replaced by optical supermodes of the compound (photonic molecule) system; and transitions between their corresponding energy levels are induced by a phonon field.  For conditions very typical of the optical microresonators studied here it behaves as a 2-level laser system, but, with a peculiar twist: the traditional roles of the material (laser medium) and cavity modes (lasing field) are reversed. Instead, the medium is now purely optical, while the laser field is provided by the material as a phonon mode. 
 \begin{figure}
\includegraphics[width=8.6cm]{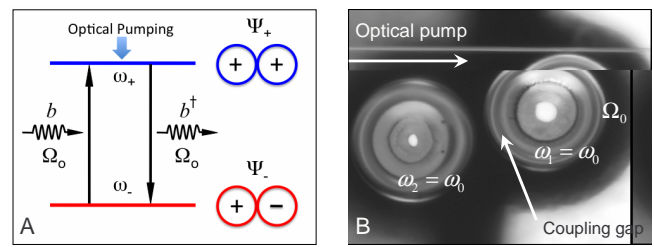}
\caption{\label{fig:tors} (A) Two level phonon laser energy level diagram including schematics of the photonic molecule symmetric and antisymmetric orbitals. (B) Physical implementation of the phonon laser using coupled microtoroids. Optical excitation and observation are enabled by a tapered fiber coupler that is visible in the image. The microtoroids are approximately 63$\mu$m in diameter: left toroid is fabricated using a 4\,$\mu$m silica layer and has an approximately 12.5\,$\mu$m minor diameter, while the right toroid is fabricated using a 2\,$\mu$m silica layer and has an approximately 8.7\,$\mu$m minor diameter.}
\end{figure}

Compound microcavity systems provide beneficial spectral controls that impact both phonon laser action and cooling; not least of which are finely-spaced optical levels whose transition energies are commensurate with phonon energies. Significantly, these level spacings are continuously tunable by adjustment of optical coupling. Amplification and cooling thus occur around a tunable linecenter;  in contrast to cavity optomechanical phenomena demonstrated to date\cite{kippenbergscience08}. Moreover, the creation of these finely spaced levels does not require increasing the microcavity dimensions, and hence has a minimal affect on the optomechanical interaction strength. Also, cooling with nearly ideal Stokes suppression and high pump coupling is possible. Other features are discussed below. 

$\it System$ $and$ $Model. \rm$ --- A schematic and micrographs of the optomechanical system are provided in figure \ref{fig:tors}.  It features two, microtoroid whispering-gallery-mode resonators \cite{armani2003}.  Although each resonator supports many optical modes, only two modes (one from each microtoroid) contribute to the physical system. These modes can be made degenerate in frequency by thermal control of the microtoroids. To couple and out-couple optical power, a tapered optical fiber is used as shown in the figure \cite{cai2000,spillane2003}.

\begin{figure}
\includegraphics[width=8.6cm]{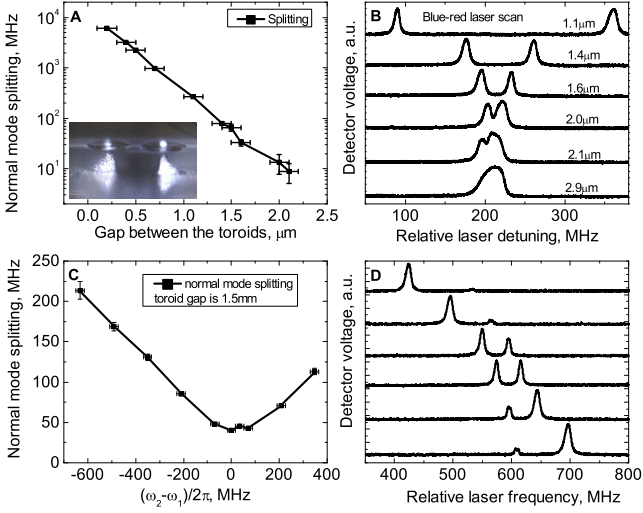}
\caption{\label{fig:splitting} Controllable splitting of the supermodes in coupled microtoroids with major diameters of 66 and 69\,$\mu$m is presented in A and B. A measured set of optical spectra taken over a range of coupling gaps appears in B, while the measured splitting frequencies as a function of gap distance are plotted in A. The inset provides a side-view of the coupled microtoroid system. Avoided crossing of the optical modes is presented in C and D. Here, the temperature of one microtoroid was gradually changed to adjust the WGM frequency. Measured spectra are presented in D and the avoided crossing as a function of frequency tuning is given in C.}
\end{figure}

Evanescent coupling of the initially uncoupled, whispering gallery modes is possible  through control of the air-gap between the microtoroids. This produces two, new normal modes (supermodes) that, for initially degenerate uncoupled modes, are symmetric and antisymmetric combinations of the uncoupled modes. Their respective eigenfrequencies are split by an amount that depends exponentially upon the air gap\cite{ilchenko94oc}.  Modal spectra at several different air gaps are presented in Fig. \ref{fig:splitting}.  These were obtained by monitoring the transmission through the fiber taper as an excitation laser is scanned in the frequency vicinity of the optical resonances. The measured, supermode splitting as a function of the air gap is also plotted to verify the exponential dependence. Control of splitting frequency from 10 MHz to nearly 10 GHz is demonstrated in the figure. Larger splittings are possible using smaller radii microtoroids so as to enhance the evanescent coupling strength.  
An alternative viewpoint of this coupling process is as an avoided crossing.  By thermal control, a microtoroid resonance at  $\omega_1$ can be scanned so as to cross  a resonance ($\omega_2$) in the other microtoroid. The normal mode splitting during such frequency scan appears as an avoided crossing, an example of which is also provided in Fig. \ref{fig:splitting}. At higher coupled powers, observation of the full spectrum was complicated by thermal nonlinearity. Modeling and observations of this nonlinearity are provided elsewhere, however, it primarily impacts pumping of the red supermode.

The Hamiltonian  for optomechanical interaction in a multimode optical cavity has been considered elsewhere  \cite{law95pra,braginsky01}.  Extension to the present system of coupled microtoroids in which one microtoroid contains a mechanical mode (frequency $\Omega_o$ and effective mass $m_{eff}$) yields, 
\begin{eqnarray}\label{eq:Hamiltonian}
H=\hbar \omega_{+}\Psi_{+}^{\dag}\Psi_{+}+ \hbar \omega_{-}\Psi_{-}^{\dag}\Psi_{-}
+\hbar \Omega_0 b^{\dag}b\\
+\frac{\hbar g x_0}{2}\left(b\Psi_{+}^{\dag}\Psi_{-}+\Psi_{-}^{\dag}\Psi_{+}b^{\dag}\right). \nonumber
\end{eqnarray}
where $\omega_+$ and $\omega_-$ are the optical frequencies of the supermodes  (the underlying, uncoupled resonances are assumed degenerate with frequency $\omega_o$), $g= {\omega_0} / {R}$ is the optomechanical coupling coefficient ($R$ is radius of the microtoroid that contains the mechanical mode), $x_0=\sqrt{{\hbar}/ {2m_{eff}\Omega_0}}$; while $\Psi_{-}$, $\Psi_{+}$ and $b$ are the lowering operators for the supermodes (defined to be red and blue, respectively) and the mechanical mode.  In deriving this Hamiltonian the energy non conserving terms as well as the conventional optomechanical terms (i.e., single mode coupling)  have been omitted.  The first interaction term in this Hamiltonian describes the destruction of one phonon and promotion of a photon from the red supermode into the blue supermode, while the second term (adjoint) describes the reverse process. The Heisenberg equations of motion for the mechanical mode and the operator $p \equiv \Psi_{-}^{\dag}\Psi_{+}$  (with damping added) are,
\begin{eqnarray}\label{eq:tevolve1}
\dot{b} = \left [-\iota \Omega_o - {\Gamma \over 2} \right ] b -\iota \frac{\Omega_R }{2} p +f(t)
\end{eqnarray}
\begin{eqnarray}\label{eq:tevolve2}
\dot{p} = \left [- \iota \Delta \omega -{\gamma \over 2} \right ] p -\iota \frac{\Omega_R }{2} \Delta N b +F(t)
\end{eqnarray}
where $\Delta \omega \equiv \omega_+ - \omega_-$; $\Delta N \equiv \Psi_{+}^{\dag}\Psi_{+} - \Psi_{-}^{\dag}\Psi_{-}$;  $\Omega_R \equiv g x_o$;  $\Gamma$ is the intrinsic energy decay rate of the mechanical mode; $2\gamma = \gamma_1 + \gamma_2$ where  $\gamma_{1,2}$ are the optical decay rates of the WGMs of each toroid; and $f(t)$, $F(t)$ are Langevin operators, included here only for completeness. The structure of these equations is equivalent to corresponding equations for a two level laser wherein $p$ is identified with the polarization of the transition, $\Delta N$ is the optical (as opposed to electronic) inversion operator, and $\gamma$ is the polarization dephasing rate (equivalently, $T_2^{-1}$).  They can be solved in the regime of weak coupling in which the optical cavity decay rate (effectively, the polarization dephasing rate) exceeds the mechanical dissipation rate (i.e., $\gamma \gg \Gamma$). Analysis, presented elsewhere, provides the mechanical gain,
\begin{eqnarray}\label{eq:gain}
G=\left(\frac{\Omega_R}{2}\right)^2\frac{\Delta N \gamma}{(\Delta \omega - \Omega_o)^2 +\left(\frac{\gamma}{2}\right)^2}. 
\end{eqnarray}
In addition to being proportional to the inversion $\Delta N$, the gain has a spectral shape that is Lorentzian with linewidth $\gamma$. Significantly, the gain spectrum line center, i.e. $\Delta \omega =\Omega_0$, can be controlled through adjustment of $\Delta \omega$ as illustrated in figure 2. Setting $\Gamma=G$, the threshold condition, gives the threshold pump power for the phonon laser as ${P}_{cT} \approx {\Gamma \gamma^2  \hbar \omega_+ / \Omega_R^2}$ (for blue-supermode excitation at linecenter). 

At a fundamental level, the scattering of the blue supermode pump photon resembles Mandelstam-Brillouin scattering\cite{braginsky01}, which, even more generally, resembles parametric down conversion\cite{Bloembergen}.  The latter viewpoint is particularly revealing as the less strongly damped field in a parametrically partnered pair will always experience stimulated emission and amplification\cite{Bloembergen}. In contrast to prior Brillouin or Raman systems that are typically restricted to the stimulated photon branch, the present system provides access to both phonon ($\gamma \gg \Gamma$ regime of this paper) and photon ($\gamma \ll \Gamma$) stimulated regimes. It is interesting to note that the above threshold power also applies for the complementary photon laser regime. While analogous in operation to recent demonstrations of Brillouin optical lasers\cite{Grudinin, Carmon}, the underlying mechanical interaction in these optomechanical photon lasers is obviously different as would be their operational features (e.g., Stokes shift).  

The delineation of the parametric instability\cite{braginsky01} into these phonon and photon stimulated regimes has not previously been noted, but is, nonetheless, crucial to microdevice applications.  Such features as the operative degree of freedom (optical or mechanical), fluctuations, coherence of motion, and gain bandwidth, to name a few, are significantly impacted by the regime. Further details, including the relationship of the current system to prior observation of optomechanical oscillation in singly resonant systems \cite{carmonprl05, kippenbergprl05,Roksari_OE_05}, will be given elsewhere. 

$\it Phonon$ $laser$ $operation. \rm$ --- A fiber laser from KOHERAS A/S with a kiloHertz-range linewidth at a wavelength of 1550\,nm served as the optical pump. The power transmitted through the tapered fiber coupler was detected with amplified New Focus detectors 1611 and 1817. To monitor coherent mechanical motion of the system, corresponding spectral components in the detected photocurrent are monitored on an electrical spectrum analyzer. Such components show a pronounced increase in magnitude when the threshold condition is achieved.  
\begin{figure}
\includegraphics[width=8.6cm]{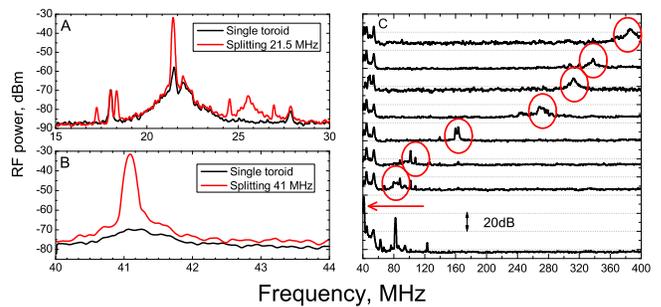}
\caption{\label{fig:lasing} A, B: Spectrum analyzer traces of the detected photocurrent. Phonon lasing spectra of two mechanical modes are shown in red. The  spectrum ``A'' corresponds to a transition frequency (i.e., supermode splitting) of 21.5\,MHz, while the spectrum ``B'' to a transition frequency  of 41\,MHz. The spectra shown in black are taken for the same optical pumping powers, but with the resonators uncoupled, thereby destroying the transition. C: Observation of amplified spectra for sub threshold pumping at a series of transition frequencies.  The gain causes a spectral bump to appear on account of amplified mechanical modes. The spectral location of the gain maximum is determined by the transition frequency (as controlled by the gap).  The transition frequency is varied from 40 MHz (lower trace) to nearly 400 MHz (upper trace). For some frequencies, the gain maximum coincides with mechanical resonances, which are subsequently ``pulled" from the noise. }
\end{figure} 

\begin{figure}
\includegraphics[width=7cm]{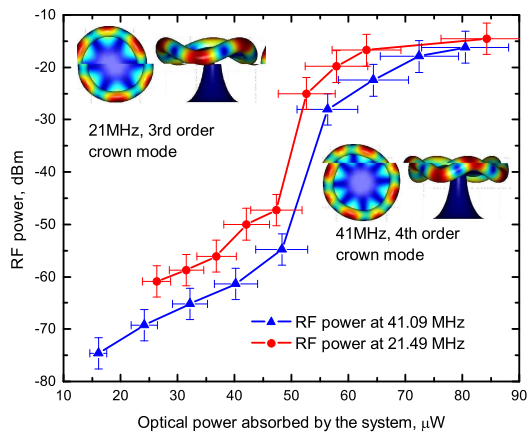}
\caption{\label{fig:ll} Plot of rf photocurrent power at frequencies indicated versus coupled optical pump power. A clear threshold is observable in the data. Also shown are the lasing mechanical ``crown" modes, inferred by matching finite element calculations with the measured oscillation frequency.}
\end{figure}

Tuning of the mechanical gain spectrum to selectively pull modes from the noise is illustrated in figure \ref{fig:lasing} (C). Each scan in this plot is taken with the air gap at a different setting so as to vary the supermode splitting and hence the mechanical gain spectrum linecenter. The overall scan frequency range is 40-400\,MHz. The ``bump" appearing at higher frequencies in the upper scan is the result of amplification. It is progressively tuned to lower frequencies by increase of the air-gap separation.  Several mechanical modes are amplified as the gain maximum comes into their spectral vicinity. Gain tuning control to produce phonon laser action is illustrated in panels A and B of figure \ref{fig:lasing},  wherein the the mechanical gain linecenter  was adjusted to coincide with mechanical modes at 21.5\,MHz and at 41\,MHz;  and pumping was sufficient to induce selective phonon laser action. Through finite element modeling, these mechanical modes were identified as third- and fourth-order crown modes, respectively, and are illustrated in figure \ref{fig:ll}. 

Log-Log plots of the RF signal power versus optical pump power confirm laser threshold behavior (Fig. \ref{fig:ll}). Both mechanical modes belong to the toroid nearest the fiber taper. Generally, modes were observed to appear in clusters containing of order 10 modes. 
This is expected because of the non-symmetrical support pillar shape caused by processing of the toroid at the edge of the wafer (so as to facilitate access to a second toroid that has been similarly processed).  When the phonon laser gain was tuned over these clusters, several mechanical modes could lase at the same time with some degree of selectivity still available. Using $m_{eff}=5\times10^{-11}$kg  \cite{kippenbergprl05}, $\Gamma=\Omega/Q_{mech}=8 \pi \times10^7/10^3=2.5\times10^5$ (estimated from mechanical spectral measurement) and $Q=2\times10^7$ (from optical spectral measurement) in the threshold equation gives an oscillation threshold of 36$\mu$W. A factor $2 \times$ is included due to excitation at full-width half maximum.  The agreement with data is reasonable in light of uncertainty in the mechanical Q due to mode clustering. 

Red-supermode pumping (negative inversion in eqn. \ref{eq:gain}) will produce damping of the mechanical mode and optomechanical cooling through absorption of mechanical quanta. Despite power limitations imposed by the thermal nonlinearity, evidence of cooling was observed as a decrease in the absolute power level of mechanical spectral peaks with increasing optical power.  Calibrated observation was not possible, however, due to the modal clustering noted above. Improvements to our processing will hopefully enable elimination of the clustering effect.

$\it Summary. \rm$ --- A tunable, phonon amplifier and phonon laser have been demonstrated that operate in close analogy to a two-level optical laser system. A compound, whispering-gallery (photonic molecule) is used to create finely spaced levels with transition frequencies commensurate to phonon frequencies; and without sacrificing cavity size or optomechanical coupling strength (i.e., g-parameter).  Continuous, wide-band tuning of the level separation, and hence the gain (and cooling) band, is possible by adjustment of resonator coupling. Also, a photon (as opposed to phonon) laser regime has been described. Evidence of mechanical cooling has been observed by excitation of the red supermode.  Moreover, for cooling, the two level system enables ideal Stokes wave suppression with excellent pump coupling. Cavities are remarkably sensitive transducers for detection of weak displacements and forces\cite{Braginsky_Measurement}; and the use of multi-resonant systems for improved readout and quantum back-action enhancement has been considered recently\cite{Kipp_BA}.   The present cavity design enables spectrally selective and tunable detection of phonons through absorption by the two level system. This detection can occur with amplification if desired. Moreover, while the embodiment described here requires additional coupling structures to link it to a phonon source, the use of optomechanical crystal based devices would enable integration of both phonon sources, detectors and vibrational waveguides  \cite{Painter_OMC}. 

$\it Acknowledgments. \rm$ --- This work was supported by DARPA.
 ISG thanks A.B. Matsko, M.L. Gorodetsky, V.B. Braginsky and Tao Lu for helpful discussions.

\end{document}